# All-optical reversible controls of integrated photonics by self-assembled azobenzene

JINGHAN HE,[1,4] ANDRE KOVACH,[2,4] DONGYU CHEN,[3] PATRICK J. G. SARIS,[1] RAYMOND YU,[3] AND ANDREA M. ARMANI[1,2,3,*]

[1]*Department of Chemistry, University of Southern California, Los Angeles, CA 90089, USA*
[2]*Mork Family Department of Chemical Engineering, University of Southern California, Los Angeles, CA 90089, USA*
[3]*Ming Hsieh Department of Electrical and Computer Engineering, University of Southern California, Los Angeles, CA 90089, USA*
[4]*These authors contributed equally to this work*
*\*armani@usc.edu*

**Abstract:** The next frontier in photonics will rely on the synergistic combination of disparate material systems. One unique organic molecule is azobenzene. This molecule can reversibly change conformations when optically excited in the blue (*trans*-to-*cis*) or mid-IR (*cis*-to-*trans*). Here, we demonstrate $SiO_2$ optical resonators modified with a monolayer of azobenzene-containing 4-(4-diethylaminophenylazo)pyridine (Aazo) with quality factors over $10^6$. Using a pair of lasers, the molecule is reversibly flipped between molecular conformations, inducing resonant wavelength shifts, and multiple switching cycles are demonstrated. The magnitude of the shift scales with the relative surface density of Aazo. The experimental data agrees with theoretical modeling.

## 1. Introduction

Optically responsive organic small molecules have played an enabling role in numerous technologies ranging from organic light-emitting diodes to organic solar cells. One particularly interesting class of organic materials are photoswitchable molecules which switch their molecular conformations upon exposure to light. Common examples include azobenzenes [1–3], stilbenes [4,5], spiropyrans [6,7] and diarylethenes [8,9]. One advantage of organic molecules in the development of active photonic devices is the ability to design a molecular structure-property relationship. Among these molecules, azobenzene is one of the most common organic photoswitches with a characteristic reversible *trans*-to-*cis* photoisomerization upon blue light exposure [10]. While this molecule has been used in photochemical information storage [11,12], functional materials [13,14], and biological detection [15–17], it has not been previously combined with integrated optical devices to create optically-triggerable, photonic platforms.

Similar to integrated electronic circuits, integrated photonic circuits are comprised of several fundamental optical building blocks including waveguides, resonators, and splitters/couplers [18–21]. From these discrete units, an optical signal can be routed on a chip, amplified in intensity or stored, converted to a different frequency, or combined with other signals. However, in order to perform these operations in a predictable manner, it is important to have dynamic control over the individual elements. Typically, this control has been achieved by using integrated electrodes to tune the refractive index of the device material via either the electro-optic [22,23] or thermo-optic [24,25] response. Unfortunately, the on-chip electrodes suffer from thermal leakage through the substrate, resulting in cross-talk between devices, ultimately, limiting device density. Photoswitchable organic materials present an alternative strategy and possible solution.

One particularly interesting integrated device is the whispering gallery mode optical resonator [26,27]. These photonic elements confine light in circular orbits at precisely defined resonant wavelengths ($\lambda$) that are determined by the optical and geometric properties of the cavity. The photon lifetime or storage time inside the cavity is defined as the quality factor (Q). Silica microtoroidal devices operating in the near-IR with ultra-high Q ($10^6$ or $10^7$) can confine light for 1,000 or 10,000 orbits for devices of diameter used in this work (50–60 μm). A small portion of the optical field forms an evanescent tail, leaking into the environment and interacting with any molecules or thin films bound to the device. The optical interaction strength between bound molecules and the circulating optical field is significantly enhanced due to the high Q as compared to a simple waveguide structure. As a result, even small changes in the index of a bound layer can change the resonant wavelength.

Previous work has investigated combining photo-active molecules and other materials with suspended microsphere optical cavities to create active or functional devices. For instance, disordered layers of biological [28,29] or azobenzene-based [30,31] photoswitchable molecules have been used to make tunable devices as well as ordered layers of nonlinear optical molecules to make frequency generators [32,33]. While innovative, the previous efforts in the field did not make the leap to an on-chip platform. Additionally, in the case of the azobenzene work, the reliance on a random or disordered deposition approach reduces the control over the magnitude of the light-molecule interaction. Covalent attachment of self-assembled monolayers of azobenzene on planar $SiO_2$ [34,35] or Au [36,37] substrates have been reported previously [38–40]. This uniform coating strategy offers a route to surface functionalize ordered azobenzene layers on optical devices.

To address these challenges, in the present work, we developed a hybrid optical device comprised of an integrated optical resonator with a grafted monolayer containing the photoswitchable azobenzene functional group in Fig. 1(a). The specific photoswitchable organic small molecule investigated was 4-(4-diethylaminophenylazo)pyridine (Aazo), and the density on the surface was varied by adding a non-photoisomerizable co-adsorbate methyl to the layer. The Aazo photoisomerizes from the thermodynamically stable *trans*-isomer to thermodynamically unstable *cis*-isomer at 410 nm and reverts to *trans*- by heating with a 10.6 μm $CO_2$ laser in Fig. 1(b,c). Because the photoswitching of the Aazo induces a refractive index change, the 1300 nm resonant wavelength is tuned, and the magnitude of the reversible frequency shift is dependent on the density of bound Aazo molecules and the intensity of the incident optical power. Complementary density functional theory (DFT) and finite element method (FEM) calculations and ellipsometry data confirm that the observed wavelength shift is due to the photoswitching of the Aazo.

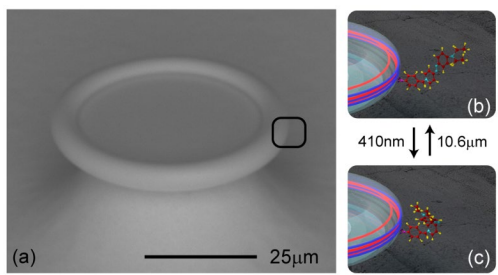

Fig. 1. (a) SEM image of Aazo-coated microtoroid resonant cavity. (b,c) Rendering of the region indicated by the box in (a) showing the reversible Aazo *trans* to (c) *cis* photoisomerization process.

## 2. Theory

The resonant cavity Q factor is defined as $Q = \lambda/\delta\lambda$ where $\delta\lambda$ is the resonance linewidth. In the present cavities, the intrinsic cavity Q is limited by the material Q factor. As shown in the expression below, in addition to being dependent on the material loss, the material Q is wavelength and refractive index dependent.

$$Q_{mat} = \frac{2\pi n_{eff}}{\lambda \alpha_{eff}} \quad (1)$$

where $n_{eff}$ is the effective refractive index and $\alpha_{eff}$ is the effective material absorption. As a result, there are two cavity Q factors of interest: the cavity Q at 410 nm ($Q_{410}$) and the cavity Q at 1300 nm ($Q_{1300}$). Additionally, each Q plays a distinctly different role in this photoswitchable, pump-probe system.

To trigger the *trans-cis* switching of the Aazo, light is coupled into a cavity resonance at 410 nm, and the optical power from this resonance initiates the photoswitching. To calculate the power circulating inside the cavity, the following general expression can be used [41]:

$$P_{circ} = P_{in} \frac{Q_o \lambda}{\pi^2 R n_{eff}} \frac{K}{(K+1)^2} \quad (2)$$

where $P_{circ}$ is the circulating power, R is the radius of the device, $n_{eff}$ is the effective index of refraction, $\lambda$ is the cavity resonant wavelength, $P_{in}$ is the input power, and $K \equiv Q_o/Q_c$ where $Q_o$ is the intrinsic Q and $Q_c$ is coupling Q (or the coupling losses of the system).

In the present system, there are two circulating powers: one at 1300 nm and the other at 410 nm. However, due to the selective absorption of the Aazo in the blue wavelength range, only the input power of the 410 nm laser will induce photo-switching. Therefore, in the Eq. 2, the values of relevance are the $P_{in}$ of the 410 nm laser and the Q and $\lambda$ at 410 nm. From this analysis, it is clear that having a higher Q in the blue wavelength range significantly reduces the amount of input power needed to induce a resonant wavelength change. In the case of the near-IR, the cavity linewidth ($\delta\lambda$) which is inversely related to the Q ($Q\sim\lambda/\delta\lambda$) determines the resolution of the resonant wavelength shift ($\Delta\lambda$). Therefore, a higher Q factor allows for smaller wavelength shifts to be detected.

The resonant wavelength position in an ultra-high Q optical cavity is defined by the cavity effective refractive index ($n_{eff}$) and R according to $\lambda = 2\pi n_{eff} R/m$, where m is the optical mode number. Therefore, changes in either parameter will induce a shift in the $\lambda$ according to $\Delta\lambda = \lambda(\Delta n_{eff}/n_{eff} + \Delta R/R)$ where $\Delta\lambda$ is the change of resonant cavity wavelength, $\Delta R$ is a change in the cavity radius, and $\Delta n$ is the change of cavity refractive index [42]. Given this dependence on R and n, the Aazo molecule studied in the present work has the potential to modify $\lambda$ through both the $\Delta R$ and the $\Delta n$ terms.

To gain insight into the relative magnitude of the $\Delta R/R$ and $\Delta n/n$ terms, DFT calculations were performed using Q-Chem 5.1 software (Q-Chem, Inc.). The gas phase ground state molecular geometry was optimized at the B3LYP/6-31G** level of theory [43,44] for two pairs of free standing Aazo molecules before calculating their dynamic polarizabilities at 1300 nm by time-dependent DFT. In this study, Aazo and [4-(chloromethyl)phenyl]-trichlorosilane (CMPS)-Aazo in both *trans-* and *cis-*isomers were considered in Fig. 2. A similar molecule, 4-[4-(N,N-dimethylamino phenyl)azo]pyridine (MAP), was used as a benchmark. MAP has the exact structure as Aazo except for two methyls on the amine group.

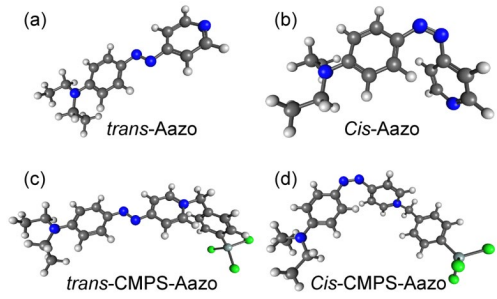

Fig. 2. The two Aazo structures in two conformations modeled using DFT: (a) *trans*-Aazo, (b) *cis*-Aazo, (c) *trans*-CMPS-Aazo, and (d) *cis*-CMPS-Aazo.

The length of each molecule was estimated by measuring the distance between the two ends of each Aazo in the optimized ground state geometry. From Table 1, the length of *trans*-Aazo and *trans*-CMPS-Aazo are 13.33 Å and 20.68 Å, respectively. The length of MAP and CMPS-MAP are previously reported as 10 Å and 18 Å, respectively [45]. This shows that the sizes of MAP and Aazo are very similar, further providing support for using MAP as the benchmarking molecule.

Because the radius of the device is unaffected by the circulating optical field at low input powers, the only contributor to the $\Delta R/R$ term is the change in the length of the molecule. Based on the calculations, as the molecule switches from *trans* to *cis*, the length change ($\Delta R$) is ~3–5 Å. As mentioned previously, the devices used in the present work have radii on the order of 20–30 μm. Therefore, the $\Delta R/R$ contribution to $\Delta \lambda$ will be approximately 1E-5.

**Table 1. Results from DFT calculations**

| Type of Aazo | Polarizability/Å$^3$ | Length/Å | N/Å$^{-3}$ | Lorenz n | $\Delta$n |
|---|---|---|---|---|---|
| *trans*-Aazo | 110.576 | 13.33 | 1/219.474 | 2.708 | 0.201 |
| *cis*-Aazo | 92.609 | 9.60 | 1/220.261 | 2.507 | |
| *trans*-CMPS-Aazo | 213.677 | 20.68 | 1/395.745 | 2.790 | 0.186 |
| *cis*-CMPS-Aazo | 182.415 | 15.26 | 1/396.532 | 2.604 | |

The index of refraction for each Aazo was estimated by combining the DFT results with the Lorenz Model [46] that describes the relationship between polarizability and index of refraction:

$$n_L = \sqrt{1 + 4\pi N p} \qquad (3)$$

where $n_L$ is the calculated refractive index of the specific Aazo, N is the average number of Aazo molecules per unit volume, and p is the mean dynamic polarizability at 1300 nm in all directions. N was calculated as the inverse of unit Aazo volume which was estimated using Connolly solvent-excluded volume. It is important to note that, because this calculation is based on solvent excluded molecules or closely packed molecules, it will over-estimate the refractive index values of the material. Thus, it should be viewed as providing an upper bound.

Based on the results calculated using Eq. (3) and summarized in Table 1, several important conclusions can be drawn. First, as the molecule photoswitches from *trans* to *cis*, the refractive index decreases, and the $\Delta$n is approximately 0.2. This decrease will induce a blueshift in $\lambda$. However, the $\Delta n/n$ term of the resonant cavity cannot be directly calculated using these values.

Since the optical field does not solely reside in the film, a calculation of $\Delta n/n$ for just the film would significantly over-estimate the impact of the film. To determine the impact of the

Δn of Aazo on the Δλ, the distribution of the optical mode across the different regions of the device (silica device, Aazo layer, and air) must be calculated. This parameter is called the effective refractive index ($n_{eff}$), and it is calculated by combining the DFT modeling of the molecule with FEM modeling of the optical mode distribution. In the present system, $n_{eff}$ is defined by:

$$n_{eff} = \alpha n_{SiO_2} + \beta n_{Aazo} + \gamma n_{air} \qquad (4)$$

where $n_{SiO2}$, $n_{Aazo}$ and $n_{air}$ are refractive index of $SiO_2$, Aazo and air, respectively. α, β and γ are effective refractive index coefficients of $SiO_2$, Aazo and air, respectively. The spatial distributions of the optical mode area of the Aazo-coated microtoroid resonators at 410 nm were analyzed using COMSOL Multiphysics FEM simulations [47,48]. This analysis accounted for the polarization state of the optical mode (TE or TM). The thickness and refractive index of the Aazo coating layer were shown in Table 1. The diameters of microtoroids were measured using scanning electron microscopy (Hitachi TM3000). The modeling results and the intensity distributions for the fundamental mode of a cavity with *trans*-CMPS-Aazo and *cis*-CMPS-Aazo is presented in Fig. 3. These results indicate that the *trans*-CMPS-Aazo has a slightly higher optical field intensity at the device surface than *cis*-CMPS-Aazo at the same mode. Additionally, if both $\Delta n_{SiO2}$ and $\Delta n_{air}$ are 0 for the *trans* to *cis* isomerization, $\Delta n_{eff}/n_{eff} = \beta\, \Delta n_{Aazo}/n_{Aazo}$.

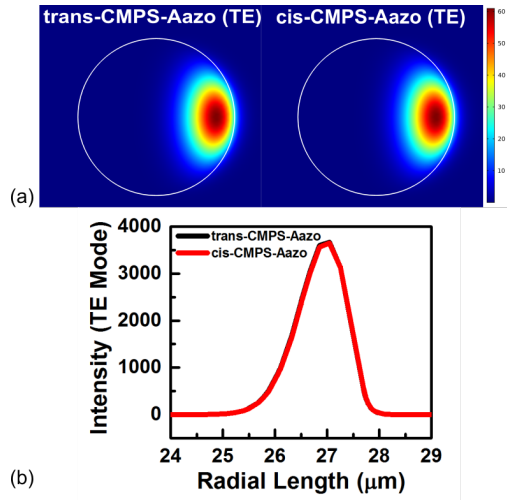

Fig. 3. FEM simulation results of *trans*-Aazo and *cis*-Aazo coated $SiO_2$ microtoroids by the COMSOL software: (a) optical mode distributions; (b) radial plots cross section plot of intensity. The difference between *trans* and *cis* is very small, and thus it is hard to distinguish.

Based on these calculations, $\Delta n_{eff}/n_{eff}$ is 4.1E-4 for the Aazo and 5.6E-4 for CMPS-Aazo. It is important to note that this calculation is based on DFT modeling of a single molecule and does not account for intramolecular interactions. However, these interactions would reduce both Δn and ΔR. Therefore, given that the $\Delta n_{eff}/n_{eff}$ term is at least an order of magnitude higher than ΔR/R (~1E-5), the approximation that $\Delta\lambda/\lambda = \Delta n_{eff}/n_{eff}$ will be made. Using these values, a theoretical upper bound on the resonant wavelength shift at 1300 nm for an entire layer of Aazo and CMPS-Aazo is estimated to be 540 pm and 730 pm, respectively.

## 3. Results and discussion

*3.1 Device Fabrication*

Si wafers (100) with 2 μm of thermal oxide ($SiO_2$) were purchased from WRS Materials. The $SiO_2$ microtoroids were fabricated according to a previously reported protocol [49]. First, using photolithography and buffer oxide etchant, the $SiO_2$ layer on the Si wafer was defined into $SiO_2$ circular pads on the surface of the wafer. Second, the $SiO_2$ disks are exposed to $XeF_2$ gas to undercut them isotropically, resulting a Si pillar underneath individual $SiO_2$ disk. Finally, the $SiO_2$ round disks are reflowed with a $CO_2$ laser (Synrad 48-2KAM) to obtain $SiO_2$ microtoroid cavities integrated on Si wafers. The major radius (R) is around 20–30 μm while the minor radius (r) is around 2.5–4.5 μm. Using these toroidal cavities as the platform device, a suite of photoswitchable Aazo monolayers are designed and anchored to the surface.

To attach the Aazo molecules to the on-chip cavities, a three-step self-assembly process is used as shown in Fig. 4(a). This silanization-based surface chemistry process results in an oriented, ordered monolayer of Aazo and was based on a previous publication with several modifications [32]. Briefly, the surface of the devices was first activated with OH groups. In the second step, the OH groups covalently bound both silanization linker agents to the surface. However, only the CMPS is reactive with the Aazo moiety. In the third step, the Aazo moiety was attached. All reagents and solvents are used as received unless otherwise noted. Aazo (98%) was purchased from Tokyo Chemical Industry. Tetrahydrofuran (THF, 99%) was purchased from VWR International. CMPS (97%) and trichloromethylsilane (TCMS, 99%) were purchased from Sigma-Aldrich.

To start with, $O_2$ plasma (SCE104, Anatech USA) was used to clean organic residues on the surface of $SiO_2$ microtoroids, yielding a monolayer of OH groups grafted on the $SiO_2$ microtoroids (Fig. 4(a)). Next, the CMPS which contains the reactive site (benzyl chloride) for the Aazo attachment was deposited on the pre-treated microtoroids. Chemical vapor deposition (CVD) for 10 min at room temperature was used to covalently bond the CMPS with the OH groups, resulting in a monolayer of CMPS on the $SiO_2$ microtoroids. Lastly, to graft Aazo on the CMPS-grafted $SiO_2$ microtoroids, Aazo in THF solution (~8.0 mM) was spin-coated on the CMPS-grafted $SiO_2$ microtoroids (7000 rpm, 30 s). To facilitate the covalent attachment between Aazo and the benzyl chloride on the CMPS-grafted $SiO_2$ microtoroids, the samples were treated in vacuum for 20 min at 120 °C. After each functionalization step, the surface of microtoroid samples were rinsed with acetone, methanol and isopropanol, and dried on a hotplate at 100 °C for 10 min. This rinsing step removed all physisorbed molecules, resulting in a covalently attached monolayer. The same procedure was used to make the control samples.

To control the surface density of the Aazo, TCMS was introduced along with the CMPS during the CVD process. The non-reactive $CH_3$ groups on the TCMS act as spacer molecules (Fig. 4(a)). Moreover, because the $CH_3$ group is relatively small, the photoisomerization of Aazo can occur unhindered. In this work, four $CH_3$:Aazo ratios as well as a control are studied: 3:1, 5:1, 7:1, 10:1, and $CH_3$ only, which are defined according to the relative ratios of TCMS:CMPS in their solutions. The actual relative ratio of $CH_3$:Aazo on devices, therefore, might be slightly different. Images of coated devices were taken by the optical microscope in Fig. 4(b), showing example images of [$CH_3$:Aazo = 10:1] and [$CH_3$ only] microtoroids. Additional control samples consisting of surface functionalized $SiO_2$/Si wafers are prepared in parallel with the optical cavities and used for surface characterization measurements.

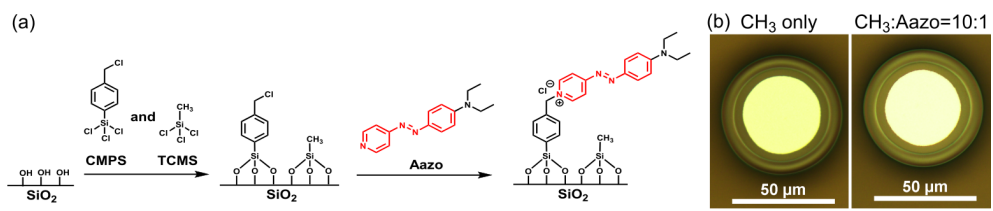

Fig. 4. The surface functionalization process of optical resonant cavities. (a) Scheme of surface functionalization of Aazo and $CH_3$ on $SiO_2$ microtoroid cavities (photoswitchable azobenzene group colored). The non-functional spacer molecule $CH_3$ is indicated. Different amounts of this molecule allowed for control of the surface density of the photoactive Aazo. (b) Representative optical microscope images of [$CH_3$ only] control and [$CH_3$:Aazo = 10:1] microtoroid cavities.

## 3.2 Characterization of photoswitchable molecules in solution

Before being anchored onto the device surface, the optical properties and switching behavior of the Aazo in THF was characterized in Fig. 5(a). UV-Vis spectra were taken by LAMBDA 950 UV/Vis Spectrophotometer (PerkinElmer). Scanning wavelength range is 300–1330 nm. For the absorption vs wavelength spectrum, *cis*-Aazo solution was acquired by irradiating initial *trans*-Aazo solution with a blue lamp (405 nm, BlueWave LED DX-1000 VisiCure, Dymax). The first scan was conducted with *trans*-Aazo which shows an absorption peak at 440 nm, an indication of $\pi$-$\pi^*$ transition. By exposing the sample to a 405 nm blue lamp for 10 min, the thermodynamically favored *trans*-isomer can be switched to the relatively unstable *cis*-isomer. The decrease in the 440 nm peak upon photoisomerization is in accordance with previously reported $\pi$-$\pi^*$ transition cases [1]. It is important to point out that both isomers of Aazo have minimal absorption at 1300 nm.

To study the photoisomerization kinetics of the *trans-to-cis* photoisomerization process, the 405 nm lamp was applied to the Aazo solution. To reverse the process, the sample was placed in a dark environment. The changes in absorbance in both directions were recorded every minute until the absorbance at 410 nm became relatively stable shown in Fig. 5(b). The change is fit to a single exponential for both photoisomerization processes, and the rate constants for *trans-to-cis* and *cis-to-trans* are $(0.5 \pm 0.1)$ min$^{-1}$ and $(0.3 \pm 0.1)$ min$^{-1}$, respectively.

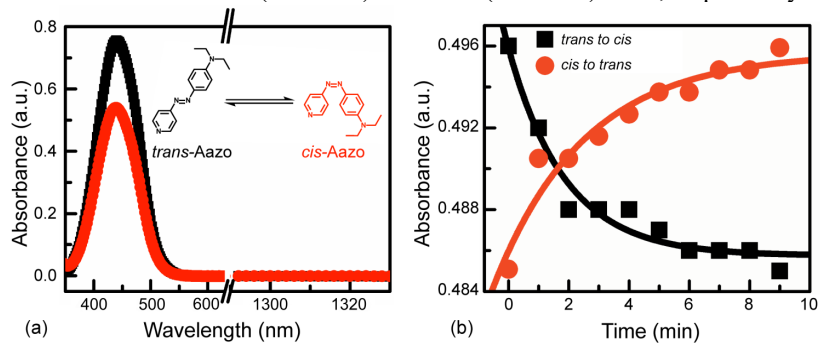

Fig. 5. UV-Vis spectroscopy results. (a) Absorption spectra of Aazo before and after photoisomerization. Inset: photoisomerization process of Aazo. (b) Kinetics study monitoring the photoisomerization progress. Excitation to induce switching occurred at 405nm and absorbance was monitored at 410nm.

## 3.3 Surface chemistry analysis

A common approach to perform elemental analysis of surfaces and to confirm surface chemistry is X-ray photoelectron spectroscopy (XPS) [50–52]. This technique was used to verify the surface attachment of the Aazo. All samples were prepared on $SiO_2$/Si wafers and stored in a desiccator before measurements. XPS data were taken from AXIS Ultra photoelectron spectrometer (Kratos Analytical Ltd) equipped with Al K$\alpha$ X-ray source (1486.6 eV). XPS measurements penetrated 10 nm below the surface of the $SiO_2/Si$ wafer samples. Survey scans were performed 5 times for each sample across 1200–0 eV with the analyzer pass energy of 160 eV. For N 1$s$, high resolution scans were conducted 15 times for each sample ranging 410–390 eV with the analyzer pass energy of 40 eV.

The XPS results in Fig. 6 verified the success of the surface functionalization of Aazo on the $SiO_2/Si$ wafers. For the initial sample, there are O 1$s$ peak at 530 eV, C 1$s$ peak at 282 eV, and Si 2$s$ peak at 152 eV. For the CMPS-functionalized sample, there is an additional Cl 2$p$ peak at 197 eV, showing the success of the CVD process. Lastly, for the Aazo-coated sample, a small N 1$s$ at 396.7 eV was confirmed by a high-resolution scan, demonstrating that Aazo is functionalized on the control substrates. The trend of peak intensities correlates to our previously reported work [32].

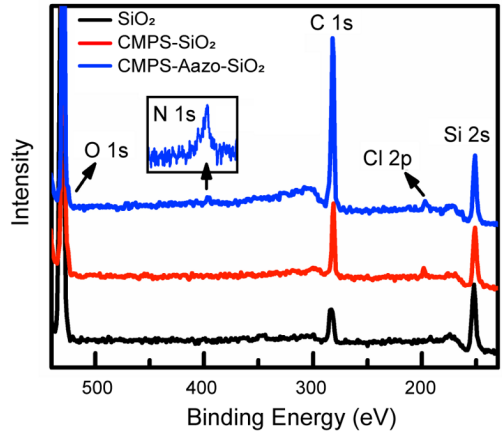

Fig. 6. XPS results verifying the two steps in the surface functionalization process. The Cl 2$p$ which is present in both CMPS samples indicates successful silanation, and the N 1$s$ indicates successful attachment of the Aazo group. The C 1$s$ is contamination in the chamber.

The measurements were conducted using a spectroscopic ellipsometer (VASE, J.A. Woollam Co.) with variable wavelengths ranging from 1000 to 1700 nm and variable incident angles ranging from 65 to 75°. The amplitude component (Φ) and phase difference (Δ) data were fit using the Sellmeier equation to determine $n_{1300}$. A 450 nm laser (~1.6 W/cm$^2$) was applied for one minute to initiate the Aazo photoisomerization on each sample. For reversing the Aazo photoisomerization, the sample was either heat-treated for 30 minutes or stored in the dark for over 24 hours. The index was measured initially, after 450nm exposure, and after treatment (thermal or dark).

The Δn induced by *trans-to-cis* isomerization ranges from 6.7E-5 for the $CH_3$ functionalized wafer to 5.7E-4 for the [$CH_3$:Aazo = 3:1] wafer. While all samples showed decreased refractive index after blue laser exposure, the $CH_3$ functionalized had a small positive index change after exposure is due to the thermo-optic effect. It is notable that these values are in excellent agreement with the results from DFT modeling. Additional discussion on these results follows in subsequent sections.

### 3.4 Device characterizations and optical Q

Given the importance of $Q_{410}$ and $Q_{1300}$ in the device performance, the device Q forms the foundation of the method. The optical testing setup used is depicted in Fig. 7(a). The cavity $Q_{1300}$ was characterized by evanescently coupling light from a 1300 nm tunable narrow linewidth laser (Velocity series, Newport) into the cavity using a tapered fiber waveguide [53]. The optical taper was fabricated by pulling a coating-stripped optical fiber (single mode fiber F-SMF-28, Newport) on a two-axis stage controller (Sigma Koki) while being heated by a hydrogen torch. The coupling distance between the taper and the microtoroid cavity was precisely controlled by a 3-axis nanopositioning stage. Photodetectors (Thorlabs) were used to detect the optical signals, displaying each signal on an oscilloscope processed with a high-speed digitizer. The transmission vs λ spectra were fit to a Lorentzian, and the loaded $Q_{1300}$ was calculated by $Q = \lambda/\Delta\lambda$, where $\Delta\lambda$ is the full width at half max. By varying the coupling strength, the intrinsic $Q_{1300}$ ($Q_{0,1300}$) was calculated using a coupled cavity modeled [54].

In parallel, a tunable 410 nm narrow linewidth laser (Velocity series, Newport) was launched into the same tapered fiber waveguide using an in-line 2:1 optical coupler (Thorlabs) with a coupling ratio of 90% (410 nm):10% (1300 nm). The loaded $Q_{410}$ ($Q_{loaded,410}$) was calculated using the same method described previously. Both lasers were tuned until cavity resonant wavelengths near 410 nm and 1300 nm were identified. It is notable that the cavity was simultaneously on-resonance at both wavelengths. Additionally, all optical cavity measurements are performed at room temperature under ambient conditions, and, when not being tested, all devices are stored under ambient conditions.To reversibly switch the Aazo between *trans* and *cis* isomerization states, first, the 410 nm laser initiated the photoswitching behavior from *trans* to *cis*. A range of 410 nm powers were studied. The resonant wavelength was tracked in real-time as soon as the 410 nm laser started to couple light into a microtoroid sample. To reverse the photoswitchable molecule from *cis* to *trans*, the 10.6 μm $CO_2$ laser (48-1KAM, Synrad) was guided by a gold coated copper mirror (Electro Optical Components, EOC) and focused by a ZnSe lens (Thorlabs) onto the microtoroid samples. The incident power from the $CO_2$ laser was approximately ~0.22 W. This wavelength is highly absorbed by silica, generating thermal energy which reverses the isomerization. However, as a result of the temperature increase, the resonant wavelength shifts due to a combination of the thermo-optic effect of silica and the molecular switching. Therefore, the cavity must thermally equilibrate in the dark to completely return to its initial position. It should be noted that the $CO_2$ laser power used to induce photoswitching is over an order of magnitude lower than those required to reflow the cavity.

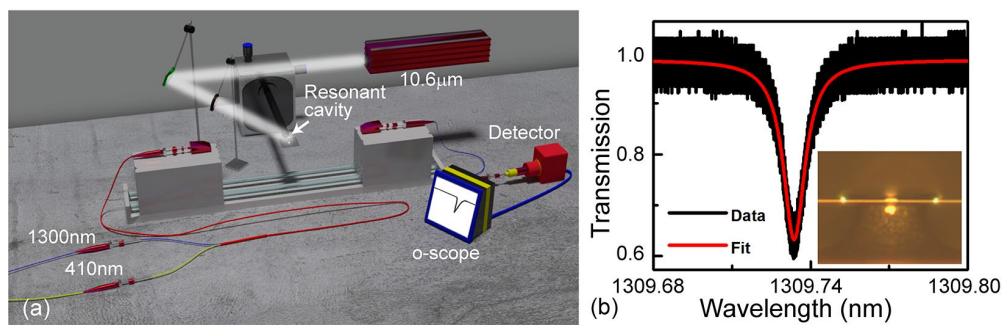

Fig. 7. Optical device characterization. (a) Optical device characterization testing setup and photoisomerization system with all components labeled. (b) Example of a Q spectrum at 1300nm. Inset is a side view optical microscope image of an Aazo-coated device coupled by an optical taper.

The tapered fiber waveguide diameter was optimized to couple light into the device at 1300 nm in Fig. 7(b). As a result, while it was possible to measure $Q_{0,1300}$, only $Q_{loaded,410}$ could be measured. Both $Q_{0,1300}$ and $Q_{load,410}$ were determined for three different devices for each coating ratio. The $Q_{0,1300}$ are plotted in Fig. 8(a), and the values range from ~$10^6$ to ~$10^7$. The $Q_{0,1300}$ values of the control device (methyl-coated) are similar to previous work using a silanization process [55]. The relatively small variations between devices is evidence of the uniformity of the surface chemistry. The $Q_{loaded,410}$ of various devices were plotted in Fig. 8(b), ranging from ~$10^5$ to ~$10^6$. For all devices in every ratio, the $Q_{loaded,410}$ is lower than the corresponding $Q_{0,1300}$.

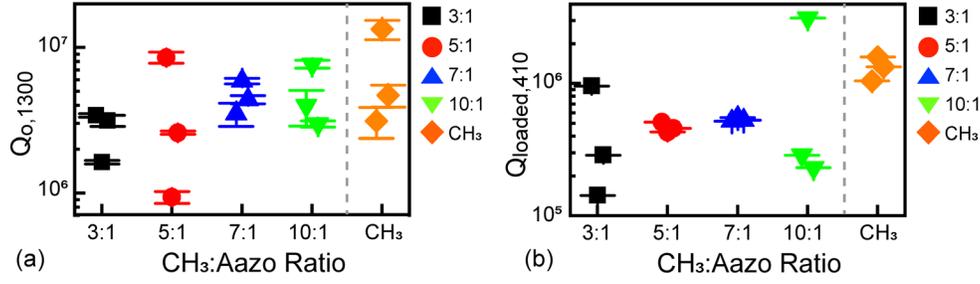

Fig. 8. Optical cavity quality factors. (a) $Q_{0,1300}$ and (b) $Q_{loaded,410}$ of Aazo-coated devices with different ratios. Each data point is a unique device.

To study the effects of photoisomerization on Q, both $Q_{0,1300}$ and $Q_{loaded,410}$ of *trans*-Aazo and *cis*-Aazo samples were measured with the same [$CH_3$:Aazo = 10:1] device (Table 2). While *trans*-Aazo has a slightly higher $Q_{0,1300}$ and $Q_{loaded,410}$ than those of *cis*-Aazo, the differences are negligible at both wavelengths; therefore, the photoisomerization process minimally affects Q.

Table 2. $Q_{0,1300}$ and $Q_{loaded,410}$ of [$CH_3$:Aazo = 10:1] device on different Aazo isomers

| Wavelength/nm | Aazo Isomer | $Q_{0,1300}$ or $Q_{loaded,410}$/$\times 10^6$ |
| --- | --- | --- |
| 1300 | *trans* | 7.68 ± 0.05 |
| 1300 | *cis* | 7.40 ± 0.03 |
| 410 | *trans* | 3.14 |
| 410 | *cis* | 2.93 |

At this point, a brief comment should be made regarding the dual wavelength tapered fiber waveguide. Using a single waveguide to simultaneously couple wavelengths into the cavity significantly reduced the experimental complexity of the system. However, because phase and index matching criteria are unable to be continuously met over the 900 nm wavelength span, it is not possible to achieve critical coupling at both wavelengths. Given the desire to balance performance, in the present work, the taper was optimized for 1300 nm while still allowing high Q modes at 410 nm to be excited. This ability is supported by the data in Fig. 8.

However, this approach clearly increased the coupling losses at 410 nm, decreasing the $Q_c$. Therefore, due to the uncertainty in coupling losses at 410 nm, the conventional approximation that K=1 should not be applied by default. However, the loaded Q values at 410 nm are less than an order of magnitude lower than the intrinsic Q values at 1300 nm. Given this data, the coupling losses cannot be significant. Therefore, in our $P_{circ}$ calculations, we are using the conventional approximation.

### 3.5 Photoisomerization-induced wavelength shifts

The entire photoswitching cycle is shown in Fig. 9(a) for the [$CH_3$:Aazo = 10:1] device with the three phases of the process indicated. First, $\lambda_{1300}$ in the device blueshifted until stabilized, for a total Δλ of approximately 7.8 pm caused by the *trans-to-cis* photoisomerization. Then, the 410 nm laser was replaced by the $CO_2$ laser (10.6 μm) which quickly heated the device within a couple of seconds. As a result, $\lambda_{1300}$ shifts due to a combination of the molecular switching and the thermo-optic effect of silica. The laser was quickly turned off, allowing the $\lambda_{1300}$ to re-equilibrate in the dark to completely return to its initial position, and the thermal energy supplied by the $CO_2$ laser returned the *cis*-Aazo back to *trans*-Aazo. The net shift for the full cycle is nearly zero, indicating the reversibility of the photoisomerization process.

To verify the reproducibility of the device fabrication process, a second device with the same CH$_3$:Aazo ratio is fabricated, and the results are shown in Fig 9(b). In this measurement, the results from multiple photoswitching events which were conducted sequentially are shown. Blueshifts and redshifts for each cycle were consistent, and the net total shift for a cycle was negligible. Due to the testing procedure used, which required measuring the power in the blue laser before and after each measurement, the noise and total P$_{circ}$ changed slightly in each cycle as indicated.

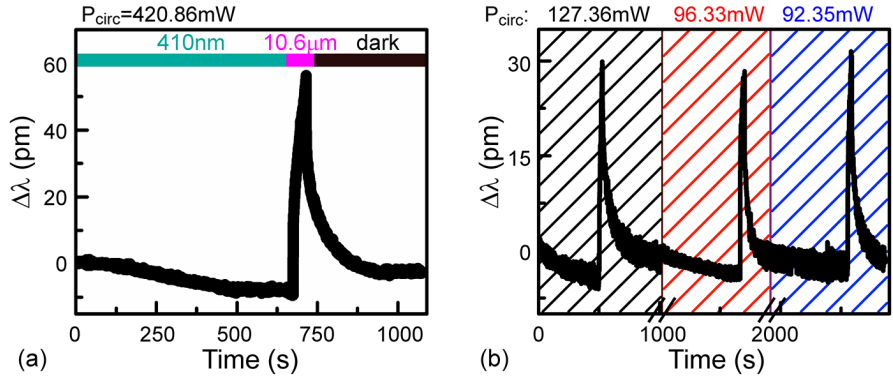

Fig. 9. Response of two different [CH$_3$:Aazo = 10:1] Aazo-functionalized optical cavities to 410nm coupled light. (a) One cycle of reversible resonant wavelength shift. When the device is exposed to the 410 nm light (P$_{circ}$ = 420.86 mW), the wavelength blue-shifted 7.8 pm. Upon exposure to the 10.6 μm wavelength, the resonant wavelength immediately red-shifted; ultimately, it returned to the initial point. (b) Repeated photo-isomerization of the Aazo is performed. The black, red and blue regions are the 1st, 2nd and 3rd cycles, respectively. The corresponding blueshifts for each cycle were 4.5 pm (P$_{circ}$ = 127.36 mW), 3.6 pm (P$_{circ}$ = 96.33 mW), and 3.0 pm (P$_{circ}$ = 92.35 mW), respectively.

To investigate the effect of the surface density of Aazo groups on the wavelength shift, similar measurements to those in Fig. 10(a) were performed with all [CH$_3$:Aazo] ratios. Additionally, to more thoroughly understand the dependence of the generated response on the circulating optical power, the amount of input blue light was changed. This data is then plotted, and the linear region is fit, allowing the optical response (Δλ/P$_{circ}$) of the device to be quantified. In performing this analysis, it is important to account for variations in device performance and size among the different cavities. For this reason, the conventional strategy is to use P$_{circ}$ in Eq. (2) instead of P$_{in}$ when comparing the optical response between devices.

A summary of all data for the *trans* to *cis* induced λ$_{1300}$ shift is shown in Fig. 10(a). The optical response was determined for each [CH$_3$:Aazo] ratio. Experimentally, the maximum power response is limited by two factors: the power output of the tunable blue laser and the range of the testing set-up over which the resonant wavelength can be monitored. Based on these results, for a given surface density, the device is operating in the linear response range, indicating that higher optical input power would continue to generate a larger response. Additionally, the optical response clearly decreases as the density of Aazo groups decreases. Because there is minimal response in the control sample, the entire wavelength change can be attributed to the Aazo group. These results are presented in terms of both Δλ and Δn (Fig. 10(a), inset).

Because the ellipsometry data was acquired with a fixed power blue laser, to directly compare the optical cavity response with the ellipsometry results, the power-dependent Δn presented in Fig. 10(a), inset must be converted to a static value. Using the previously defined expression, P$_{circ}$ is set to a fixed quantity (200 mW) allowing this value to be determined in Fig. 10(b). However, it is not expected that the two experimental data sets would match precisely because the optical field overlaps are different, but the general trend should scale. As shown in

Fig. 10(b), the ellipsometry data and the resonant cavity response is similar. This provides further evidence that the blue shift can be solely attributed to the photoswitchable molecule and, by varying the density of molecules on the surface, the magnitude of the response can be controlled.

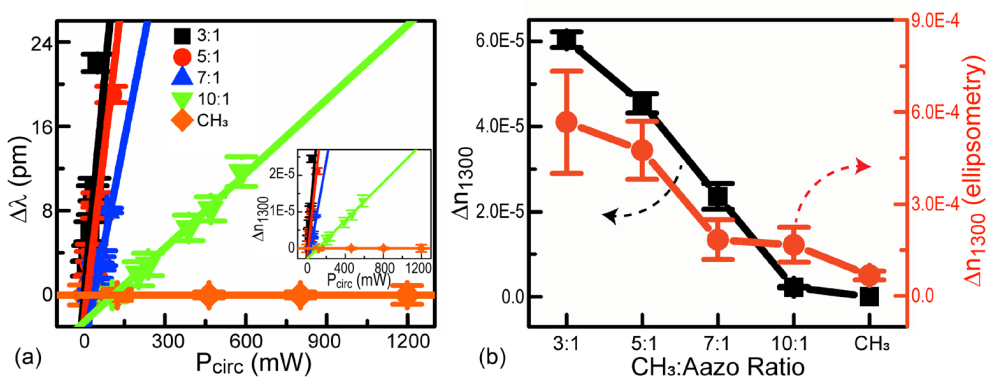

Fig. 10. Response as a function of Aazo surface concentration and input power. (a) For a given surface density of Aazo groups, as the amount of circulating power increases, the total resonant wavelength shift also increases. The slopes of these lines are the optical response of devices which are characteristic to the surface density of Aazo groups. Inset: The magnitude of the index change scales with the amount of circulating power in the device. (b) Using the optical responses from (a) and setting $P_{circ}$ = 200 mW, the resonant cavity results can be compared to the ellipsometry results. The general behavior is similar.

## 4. Conclusions

By combining photoswitchable organic molecules with integrated optical devices, all-optical control of the resonant wavelength of integrated photonic resonant cavities is demonstrated. To enhance the performance, a single waveguide was used to simultaneously excite the visible resonance used for tuning the near-IR resonant modes. The blue laser photoswitches the *trans*-Aazo into the *cis*-Aazo, resulting in a reduction of the refractive index and a blue-shift in the resonant wavelength. This photoisomerization can be recovered by exposing the device to a $CO_2$ laser. The optical response of laser-induced photoswitching of Aazo induces a cavity index change on Aazo-coated devices which is associated with the surface density of Aazo groups and the circulating power of the 410 nm laser. The trend of ellipsometry measured index change correlates to that of the cavity index change on each [$CH_3$:Aazo] ratio. This approach may provide a new strategy for all-optical encoding the resonant wavelength of integrated photonic devices, which could improve the cost-performances compared to electronic encoding techniques. On the other hand, tracking the kinetics studies in photochemical reactions with such an approach may also be promising.

## Funding

Army Research Office (ARO) (W911NF1810033) and Office of Naval Research (ONR) (N00014-17-2270).

## Acknowledgments

The authors would like to thank Mark Veksler (University of Southern California) for renderings.

## Disclosures

The authors declare no conflicts of interest.